\documentclass[12pt]{article}

\usepackage{amsmath}
\usepackage{amsfonts}
\usepackage{amssymb}
\usepackage{multirow}

\def \n{\noindent}

\newtheorem{Theo}{Theorem}

\newtheorem{Prop}{Proposition}

\begin{document}

\begin{titlepage}
\begin{center}
        \vspace*{1cm}

        \raggedright
        \begin{center}
        \LARGE
        \textbf{Empirical Lossless Compression Bound of a Data Sequence}
        \end{center}

        \vspace{1.5cm}
        \large
        \textbf{Author:} {\normalsize
        Lei M Li\,$^{\text{1,2,}*}$}

        \vspace{0.8cm}
        \large
        \textbf{Affiliations: \\}
        \normalsize
        $^{\text{1}}$Academy of Mathematics and Systems Science, Chinese Academy of Sciences, Beijing 100190, China. \\
        $^{\text{2}}$University of Chinese Academy of Sciences, Beijing 100049, China\\

        \vspace{0.8cm}
        $^\ast$ Correspondence should be addressed to Lei M Li (lilei@amss.ac.cn)\\
        Telephone: +8610--82541585\\
        Fax: +8610--6265--8364\\

        \vspace{0.8cm}
        \vspace{3cm}
    \end{center}
\end{titlepage}
\newpage

\begin{abstract}
We consider the lossless compression bound of any individual data sequence.
Conceptually, its Kolmogorov complexity is such a bound yet uncomputable.
The Shannon source coding theorem states that the average
compression bound is $nH$, where $n$ is the number of
words and $H$ is the entropy of an oracle probability distribution
characterizing the data source.
The quantity $nH({\hat \theta}_n)$ obtained by
plugging in the maximum likelihood estimate is an underestimate of
the bound.
Shtarkov showed that the  normalized maximum likelihood (NML)
distribution or code length is optimal in a minimax sense for any parametric family.
In this article, we consider the exponential families, the only models
that admit sufficient statistics whose dimensions remain bounded as the sample size grows.
We show by the local asymptotic normality that the NML code length is
$nH(\hat \theta_n)
+\frac{d}{2}\log \, \frac{n}{2\pi}
+\log \int_{\Theta} |I(\theta)|^{1/2}\, d\theta+o(1)$, where $d$ is
the model dimension or dictionary size, and $|I(\theta)|$ is the determinant of the Fisher information matrix.
We also demonstrate that  sequentially predicting the optimal code
length for the next word via a Bayesian mechanism leads to the mixture code,
whose pathwise length is given by
$nH({\hat \theta}_n)
+\frac{d}{2}\log \, \frac{n}{2\pi}
+\log \frac{|\, I({\hat \theta}_n)|^{1/2}}{w({\hat \theta}_n)}+o(1)
$, where $w(\theta)$ is a prior.
If we take the Jeffreys prior when it is proper, the expression agrees with the
NML code length.
The asymptotics apply to not only discrete symbols but also continuous data
if the code length for the former is replaced by the description length for the latter.
The analytical result is exemplified by calculating
compression bounds of protein-encoding DNA sequences
under different parsing models.
Typically, the highest compression is achieved when the parsing is
in the phase of the amino acid codons. On the other hand, the compression
rates of pseudo-random sequences are larger than 1 regardless of parsing models. These model-based results are
consistent with that random sequences are incompressible as asserted
by the Kolmogorov complexity theory.
The empirical lossless compression bound is particularly more accurate when the dictionary size is relatively large.
\end{abstract}

\section {Introduction}\label{Introduction}
The computation of the compression bound of any individual sequence is both a philosophical and a practical problem.
It touches on the fundamentals of human being's intelligence.
After several decades of efforts, many insights have been gained by experts from different disciplines.

In essence, the bound is the shortest program that prints the sequence on a Turing machine, referred to as the Solomonoff-Kolmogorov-Chaitin
algorithmic complexity.
Under this setting, if
a sequence cannot be compressed by any computer program, it is random.
On the other hand,
if we can compress the sequence by a certain program or coding
scheme, it is then not random, and we learn some pattern or knowledge in the sequence.
Nevertheless, this Kolmogorov complexity is not computable.

Along another line,
the source coding theorem proposed by Shannon \cite{Shannon-original} claimed that the optimal coding, or the average shortest code length, is no less than $n H$, where $n$ is the number of
words and $H$ is the the entropy of
the source if its distribution can be specified.
Although Shannon's probability framework has inspired the inventions of some ingenious
compression methods, $n H$ is an oracle  bound.
Some further questions need to be addressed.
First, where does the probability distribution come from?
A straightforward solution is one inferred from the data themselves.
However, in the case of discrete symbols, plugging in the word
frequencies $\hat \theta_n$ observed in the sequence results in $nH(\hat \theta_n)$,
which is, as can be shown, an underestimate of
the bound.
Second, the word frequencies are counted according to a dictionary. Different dictionaries
would lead to different distributions or codes. What is the criterion for selecting a good dictionary?
Third, the practice of some compression algorithms such as the Lempel-Ziv coding shows
as the length of a sequence gets longer, the size of the dictionary gets larger.
What is the exact impact of the dictionary size on the compression?
Fourth, can we achieve the compression limit by a  predictive code that goes through data
in only one round?
Fifth, how is the bound derived from the probability framework,
if possibly, connected to the conclusions drawn from the algorithmic complexity?

In this article, we review the key ideas of lossless compression
and present some new mathematical results relevant to
the above problems. Besides the algorithm complexity and the
Shannon source coding theorem, the technical tools
center around
the normalized  maximum likelihood (NML) coding \cite{Shtarkov-1987,930912} and predictive coding
\cite{Clarke-Barron-jeffrey,my-2000-ieee}.
The expansions of these code lengths lead to
an empirical compression bound, which is indeed sequence-specific and
thus has a natural link to algorithmic complexity.
Although the primary theme is the pathwise asymptotics, their related average results were
discussed as well for the sake of comparison.
The analytical results apply to not only discrete symbols but also continuous data
if the codelength for the former is replaced by the description length for the latter \cite{Barron-R-Yu-98}.
Other than theoretical justification,
the empirical bound is exemplified by  protein-coding
DNA sequences and pseudo-random sequences.
\section {A brief review of the key concepts}\label{review}
\paragraph {Data compression}\label{Data compression}
The basic concepts of lossless coding can be found in
the textbook  \cite{Cover-book}.
Before we proceed, it is helpful to clarify some
jargon used in this paper: strings, symbols, and words.
We illustrate them by an example. The following
``studydnasequencefromthedatacompressionpointofviewforexampleabcdefghijklmnopqrstuvwxyz'',
is a  string. The 26 small case distinct English letters
appearing in the string
are called
symbols, and they form an alphabet.
If we parse the string into
``study'', ``dnasequence'', ``fromthedata'', ``compressionpointofview'', ``forexample'',
``abcdefg'', ``hijklmnopq'', ``rstuvwxyz'', and
these substrings are called words.

The implementation of data compression includes an encoder and a decoder.
The encoder parses the string to be compressed into words and replaces each word by its codeword. Consequently, this produces
a new string, which is hopefully shorter than the original one
in terms of its.
The decoder, conversely, parses the new string into codewords, and interpret
each codeword back to a word of the original symbols.
The collection of all distinct words in a parsing is called a dictionary.

In the notion of data compression, two issues arise naturally. First,
is there a lower bound?
Second, how do we compute this bound, or
is it computable at all?
\paragraph {Prefix code}\label{Prefix code}
A basic idea in lossless compression is the prefix code or instantaneous code.
A code is called a prefix one if no
codeword is a prefix of any other codeword.
The prefix constraint
has a very close relationship to the metaphor of the
Turing machine, by which the algorithmic complexity is defined.
Given a prefix code over an alphabet of $\alpha$ symbols,
the codeword lengths $l_1$, $l_2$, $\cdots$, $l_m$,  where
$m$ is the dictionary size, must
satisfy the  Kraft inequality: $\sum_{i=1}^m \, \alpha^{-l_i}\le 1$.
Conversely, given a set of code lengths that satisfy this
inequality, there exists a prefix code with these
code lengths.
Please notice that the dictionary size in a prefix code could be either
finite or countably infinite.

The class of prefix codes is smaller than
the more general class of uniquely decodable
codes, and one may expect
that some uniquely decodable codes
could be advantageous over  prefix codes in terms of
data compression.
However, this is not exactly true, for it can be shown that the codeword lengths of any
uniquely decodable code must satisfy the Kraft inequality.
Thus we can construct a prefix code to match the codeword
lengths of any given uniquely decodable code.

A  prefix code has an attractive self-punctuating feature: it can be decoded without reference to
the future codewords, since the end of a codeword is immediately
recognizable.
For these reasons, people stick to prefix coding
in practice.
A  conceptual yet convenient generalization of the Kraft inequality is to
drop the integer requirement for code lengths and ignore the effect of rounding up.
A general set of code lengths can be implemented by the arithmetic coding \cite{Rissanen-1976-arithmetic-code,Rissanen-Langdon-1979}.
This generalization leads to a correspondence between
probability distributions and prefix code lengths:
to every distribution $P$ on the dictionary, there
exists a prefix code $C$
whose length $L_{C}(x)$ is
equal to $-\log P(x)$ for all words $x$.
Conversely, for every prefix code $C$ on the dictionary, there
exists a probability measure $P$ such that $-\log P(x)$ is equal to the code length $L_C(x)$ for all words $x$.
\paragraph{Shannon's probability-based coding}\label{Shannon theory}
In his seminal work \cite{Shannon-original},
Shannon proposed the  source coding theorem based on
a probability framework.
If we assume a finite number of words $A_1$, $A_2$, $\cdots$,
$A_m$ are generated from a
probabilistic source denoted by a random
variable $X$ with frequencies $p_i$, $i=1, \cdots,m$,
then the expected length of any prefix code
is no shorter than the entropy of
this source defined as: $H(X)=-\sum_{i=1}^m \, p_i \log p_i$.
Throughout this paper,  we take 2 as the base of the logarithm operation,
and thereby bit is the unit of code lengths.
This result offers  a lower bound of data compression if
a  probabilistic model can be assumed.

Huffman code is such an optimal code that reaches the expected
code length. The codewords are defined by a binary tree built from word frequencies.
Shannon-Fano-Elias code is another one that uses at most two bits more
than the lower bound. The code length
of $A_i$ in Shannon-
Fano-Elias code is approximately equal to
$-\log p_i$.
\paragraph{Kolmogorov  complexity and algorithm-based coding}
\label{Kolmogorov algorithmic complexity}
Kolmogorov, who laid  out the foundation of the probability theory,
interestingly put away probabilistic models,
and along with other researchers including Solomonoff and Chaitin,
pursued another path to understand the information structure
of data based on the notion of the universal Turing machine.
Kolmogorov \cite{10.1214/aop/1176991250} expressed the following: ``information theory must precede probability
theory, and not be based on it.''

We give a brief account of some facts about Kolmogorov complexity
relevant to our study, and refer readers to Li and Vit\'{a}nyi \cite{Li-Vat-book},
Vit\'{a}nyi and  Li \cite{Li-Vat-2000-ieee} for detail.
A Turing machine is a computer with a finite state
operating on a finite symbol set, and
is  essentially the abstraction of
any concrete computer that has CPUs, memory, and
input and output devices.
At each unit of time, the machine reads in one
operation command from the program tape, write
some symbols on a work tape, and change its state according
to a transition table.
Two important features need more explanation.
First,  the program is linear,  namely, the machine reads the tape from
left to right, never goes back.
Second, the program is prefix-free, namely, no program
leading to a halting computation can be the prefix of
another such program. This feature is an analog to
the prefix-coding idea.
A universal Turing machine can reproduce the results of other machines.
Kolmogorov complexity of a word $x$ with respect
to a universal computer $\mathcal {U}$, denoted
by $K_{\mathcal {U}}(x)$, is defined
as the minimum length overall programs that print
$x$ and halt.

The Kolmogorov complexities of all words  satisfy
the Kraft inequality due to its natural connection to prefix coding.
In fact, for a fixed machine $\mathcal {U}$,
we can encode $x$ by the minimum length program
that prints $x$ and halt.
Given a long string, if we define a way to parse
it into words, then we encode each word by the above
program. Consequently, we encode the string by concatenating
the programs one after another. The decoding can easily be
carried out by inputting the concatenated program into
$\mathcal {U}$.
One obvious way of parsing is to take the string
itself as the only word. Thus how much we can
compress the string depends on the complexity of
this string. At this point,
we see the connection between
data compression and the Kolmogorov complexity, which
is defined for each string
on an implementable type of computational machine --- the Turing machine.

Next, we highlight some theoretical results about Kolmogorov complexity.
First, it is not machine specific except for a machine-specific constant.
Second, the Kolmogorov complexity is unfortunately
 not computable.
Third, there exists a universal probability
$P_{\mathcal{ U}} (x)$ with respect to a universal machine,
such that $2^{-K(x)}\le P_{\mathcal{ U}} (x) \le c\,2^{-K(x)}$
for all strings, where $c$ is a constant independent of $x$.
This means that $K(x)$ is equivalent to
$-\log P_{\mathcal{ U}} (x)$ except for a constant, which can be viewed
as the code lengths of a prefix code in light of the
Shannon-Fano-Elias code.
Because of the non-computability of Kolmogorov complexity,
the universal probability is not computable either.

The study of the Kolmogorov complexity tells us
that the assessment of exact compression bounds of strings
is beyond the ability of any specific Turing machine. However,
any program on a Turing Machine offers, except for a constant, an upper bound.
\paragraph{Correspondence between probability models and string parsing}
A critical question remained to be answered in the Shannon source coding theorem is: where does the model that
defines probabilities come from?
According to the theorem, the optimal code lengths are proportional to the
negative logarithm of the word frequencies.
Once the
dictionary is defined, the word frequencies can be counted for any individual
string to be compressed.
Equivalently, a dictionary can be induced by the way we parse a string.
It is noted that the term ``letter'' instead of ``word'' was used in Shannon's original paper
\cite{Shannon-original}, which did not discuss how to parse strings into words at all.
\paragraph{Fixed-length and variable-length parsing}\label{parsing}
The words generated from the parsing process could be either of
the same length or of variable lengths. For example,
we can encode Shakespeare's work letter by letter, or
encode them by natural words of different lengths.
A choice made at this point leads to two quite
different coding schemes.

If we decompose a  string into words of the same number of
symbols, this is a fixed-length parsing.
The two extra bits for each word
is a big deal if the number of symbols in each word is small.
As the word length gets longer and longer,
the two extra bits are relatively negligible for each
block.
An effective alternative to get around the issue of extra bits is the
arithmetic coding  that integrates the codes of successive words
at the cost of more computations.

Variable-length parsing decomposes a  string into words of
variable number of
symbols.
The popular Lempel-Ziv coding is such a scheme.
Although the complexity of a string $x$ is not computable,
the complexity of '$x\:1$',  relative to '$x$' is small.
To concatenate an '$1$'
to the end of '$x$', we can simply use the program of printing $x$ followed by
printing '1'. A recursive implementation of this idea leads to the
Lempel-Ziv coding, which concatenates the address of '$x$' and
the code of '1'.

Please notice that  as the data length increases, the dictionary size resulted from
the parsing scheme of the Lempel-Ziv coding
increases as well if we do not impose an upper limit.
Along the process of encoding, each word occurs only once because down the road either it will  not be a prefix of any other word
or a new word concatenating it with a certain suffix symbol will be found.
To a good approximation, all the words encountered up to a point are equally likely.
If we use the same bits to store the addresses of these words, their code lengths are equal.
Approximately, it obeys
Shannon's source coding theorem too.
\paragraph{Parametric models and complexity}\label{Models of finite dimensions}
Hereafter we use parametric probabilistic models to
count prefix code lengths. The specification of a parametric model includes three aspects:
a model class; a model dimension;
and parameter values. Suppose we restrict our attention to
some hypothetical model classes. Each of these model classes
is indexed by a set of parameters, and we call the number of parameters in
each model its dimension.
We also assume the identifiability of the parameterization, that is,
different parameter values correspond to different models.
Let us denote one such model class by a probability measure
$\{P_{\theta}: \theta \in \mbox{an open set } \Theta \subset {\bf R}^d \}$, and
their corresponding frequency functions by $\{p(x;\theta)\}$.
The model class is usually defined by a parsing scheme.
For example, if we parse a string symbol by symbol, then
the number of words equals the number of symbols appearing
in the string. We denote the number of symbols by $\alpha$, then $d=\alpha-1$. If we parse the string by every two symbols,
then the number of words increases to $d=\alpha^2-1$, and so on.

From the above review of  Kolmogorov complexity, it is clear
that strings themselves do not admit probability models in the first place.
Nevertheless, we can fit a string by a parametric model. By doing so,
we need to pay extra bits
to describe the model, as observed by Dr. Rissanen. He termed them as stochastic complexity or parametric complexity.
The total code lengths by a model include both the data description and the parametric complexity.

\paragraph{Two references for code length evaluation}
The evaluation of redundancy of a given code needs a reference.
Two such references are discussed in the literature. In the first scenario,
we assume that the words $X^{(n)}=\{X_1, X_2\cdots, X_n\}$ are generated
according to $P_{\theta_0}$ as
independent and identically
distributed (iid) random variables, whose outcomes are denoted by $\{x_i\}$.
Then the optimal code length is given by
$L_{0}=-\sum_{i=1}^{n}\, \log p(X_i;\theta_0)$.
As $n$ goes large, its average code length is given by
$EL_{0}=nH(\theta)$.
In general, the code length corresponding to any distribution $Q(x)$ is given by
$L_{Q}=-\sum_{i=1}^{n}\, \log q(X_i)$, and its redundancy is
$R_{Q}=L_{Q}-L_{0}$. The expected redundancy is the Kullback-Leibler divergence between the two distributions:
\begin{equation*}
E_{P_{\theta_0}}(L_{Q}-L_{0})=E_{P_{\theta_0}}\log \frac{P_{\theta_0}(X^{(n)})}{Q(X^{(n)})}=D(P_{\theta_0}||Q) \ge 0\, .
\end{equation*}
It can be shown that minmax
and maxmin values of average redundancy are equal \cite{Haussler-minimax}.
\begin{equation*}\label{maxmin=minmax}
\inf_Q \sup_{\theta}E_{P_{\theta}}\log \frac{P_{\theta}(X^{(n)})}{Q(X^{(n)})}=\sup_{\theta}\inf_Q E_{P_{\theta}}\log \frac{P_{\theta}(X^{(n)})}{Q(X^{(n)})}
=I(\Theta;X^{(n)})\, .
\end{equation*}
Historically a key progress on redundancy \cite{Krichevskii_1968,Rissanen-86-annals}
is that
for each positive number $\epsilon$ and for all $\theta_0 \in \Theta$ except in a set whose volume goes to zero,
as $n\longrightarrow \infty$,
\begin{equation}\label{rissanen-result}
E_{P_{\theta_0}}(L_{Q}-L_{0})\ge \frac{d-\epsilon}{2}\log \, n.
\end{equation}
All these results are about average code length over all possible strings.

Another reference which any code can be compared with  is
obtained by replacing $\theta_0$ by the  maximized likelihood estimate
 $\hat \theta_n$ in $L_{0}$, that is,
$L_{\hat \theta_n}=-\sum_{i=1}^{n}\, \log p(X_i;\hat \theta_n)$. Please notice that
$L_{\hat \theta_n}$ does not satisfy the Kraft inequality.
This perspective is a practical one, since in reality $x^{(n)}$ is simply data without
any probability measure. Given a parametric model class $\{P_{\theta}\}$, we fit the data
by one surrogate model that maximizes the likelihood.
Then we consider
\[
L_{Q}-L_{\hat \theta_n}=\log \frac{p(x^{(n)};\hat \theta(x^{(n)}))}{q(x^{(n)})}\, .
\]
\paragraph{Optimality of normalized maximum likelihood code length}\label{NML_optimality}
Minimizing the above quantity leads to the normalized maximum-likelihood  (NML) distribution:
\[
{\hat p}(x^{(n)})=\frac{p(x^{(n)};\hat \theta(x^{(n)}))}{\sum_{x^{(n)}}\,p(x^{(n)};\hat \theta(x^{(n)}))}\, .
\]
The NML code length is thus given by
\begin{equation}\label{empirical-likelihood-logsum}
L_{NML}=-\log\, p(x^{(n)};\hat \theta(x^{(n)}))+\log\,{\sum_{x^{(n)}}\,p(x^{(n)};\hat \theta(x^{(n)}))}\, .
\end{equation}
Shtarkov \cite{Shtarkov-1987} proved the optimality of NML code by showing it solves
\begin{equation}\label{nml-optimality}
\min_{q}\max_{x^{(n)}}\log \frac{p(x^{(n)};\hat \theta(x^{(n)}))}{q(x^{(n)})}\, ,
\end{equation}
where $q$ ranges over the set of virtually all distributions.
Later Rissanen \cite{930912} further proved that NML code solves
\begin{equation*}
\min_{q}\max_{g}E_{g}[\log \frac{p(X^{(n)};\hat \theta(X^{(n)}))}{q(X^{(n)})}]\, ,
\end{equation*}
where $q$ and $g$ range over the set of virtually all distributions. This result states that the NML
code is still optimal even if the data are generated from outside the parametric model family.
Namely, regardless of the source nature in practice, we can always find the optimal code length
from a distribution family.
\section{Empirical code lengths based on exponential family distributions}\label{Empirical lossless coding length}
In this section, we fit the data from a source, either discrete or continuous, by an exponential family due to
the following considerations. First, the multinomial distribution, which is used to encode discrete symbols,
is an exponential family.
Second, according to the Pitman-Koopman-Darmois theorem, exponential families are,
under certain regularity conditions, the only models
that admit sufficient statistics whose dimensions remain bounded as the sample size grows.
On one hand, this property is most desirable in data compression. On the other hand,
the results would be valid in the more general
statistical learning other than source coding.
Third, as we will show, the first term in the code length expansion  is
nothing but the empirical entropy for exponential families, which is a straightforward extension of
Shannon's source coding theorem.
\paragraph{Exponential families}\label{exponential_family}
Consider a canonical
exponential family of distributions
$\{P_{\theta}: \theta \in \Theta\}$, where the natural parameter space
$\Theta$ is an open set of ${\bf R}^d$.
The density function is given by
\begin{equation}\label{exponential}
p(x;\theta)=\exp\{\theta^{T}\,S(x)-A(\theta)\}\, ,
\end{equation}
with respect to some measure $\mu(dx)$ on the support of data.
The transposition
of a matrix (or vector) $V$ is represented by $V^{T}$ here and throughout the paper.
$S(\cdot)$ is the sufficient statistic for the parameter $\theta$.
We denote the first and the second
derivative of $A(\theta)$  respectively by $\dot {A}(\theta)$ and
$\ddot {A}(\theta)$.
The entropy or differential entropy of $P_\theta$ is:
$H(\theta)=A(\theta)-\theta^{T}\dot A(\theta)$.
The following result is an empirical and pathwise version of Shannon's source coding theorem.
\begin{Theo}{\bf (Empirical optimal source code length)}\label{main-theorem}
If we fit an individual data sequence by an exponential family distribution, the NML code length is given by
\begin{equation}\label{NML-codelength}
{L_{NML}}=nH(\hat \theta_n)
+\frac{d}{2}\log \, \frac{n}{2\pi}
+\log \int_{\Theta} |I(\theta)|^{1/2}\, d\theta+o(1)\, ,
\end{equation}
where $H(\hat \theta_n)$ is the entropy evaluated at the maximum likelihood estimate (MLE)
$\hat \theta_n=\hat \theta(x^{(n)})$, and $|I(\theta)|$ is the determinant of the
Fisher information $I(\theta)=[-E(\frac{\partial ^2{log p(X; \theta)}}{{\partial \theta_j}{\partial \theta_k}})]_{j,k=1, \cdots, d}$.
The integral in the expression is assumed to be finite.
\end{Theo}
The first term in \eqref{empirical-likelihood-logsum} is
$n A(\hat{\theta}_n)-[\sum_{i=1}^n S(x_i)]^{T} {\hat{\theta}_n}=n A(\hat{\theta}_n)-n\dot {A}({\hat{\theta}_n})^{T} {\hat{\theta}_n}=nH({\hat{\theta}_n})
$,
namely, the entropy in Shannon's theorem except that
the model parameter is replaced by the MLE.
The second term has a close relationship
to the BIC work of Akaike \cite{Akaike-BIC-77} and Schwartz \cite{schwarz-BIC}, and the third term is the Fisher information which
characterizes the local property of a distribution family.
Surprisingly and interestingly, this empirical version of the lossless coding theorem puts together
the three pieces of fundamental works respectively by Shannon, Akaike-Schwartz, and Fisher.

Next, we give a heuristic proof of \eqref{NML-codelength} by the local asymptotic normality (LAN)
\cite{LeCam-Yang-book}, though a complete proof can be found in the Appendix.
In the definition of NML code length (\ref{empirical-likelihood-logsum}), the first term
becomes empirical entropy for exponential families. Namely,
\begin{equation}\label{entropy-logsum}
{L_{NML}}=nH(\hat \theta_n)
+\log \, {\sum_{x^{(n)}}\,p(x^{(n)};\hat \theta_n)}\, .
\end{equation}
The remaining difficulty is the computation of the summation.
In a general problem of data description length, Rissanen \cite{Rissanen-96}
derived an analytical expansion requiring five assumptions, which were hard to verify.
Here we show for sources from exponential families, the expansion is valid as long as
the integral is finite.

Let $U(\theta, \frac{r}{\sqrt{n}})$ be a cube of size $\frac{r}{\sqrt{n}}$ centering at $\theta$, where $r$ is a constant.
LAN states that we can expand probability density in each neighborhood $U(\theta, \frac{r}{\sqrt{n}})$ as follows.
\[
\log \frac{p(x^{(n)};\theta+h)}{p(x^{(n)};\theta)}=h^{T}[\sum_{i=1}^{n}\, S(x_i)-n\dot {A}(\theta)]-\frac{1}{2}h^{T}[nI(\theta)]h+o(h)\, ,
\]
where $I(\theta)=\ddot {A}(\theta)$. Maximizing the likelihood in $U(\theta, \frac{r}{\sqrt{n}})$ with respect to $h$ leads to
\[
\max_{h}\log \frac{p(x^{(n)};\theta+h)}{p(x^{(n)};\theta)}=\frac{1}{2}[\sum_{i=1}^{n}\,S(x_i)-n\dot {A}(\theta)]^{T}[nI(\theta)]^{-1}[\sum_{i=1}^{n}\, S(x_i)-n\dot {A}(\theta)]+o(\frac{r}{\sqrt{n}})\, .
\]
Consequently, if ${\hat \theta}_n(x^{(n)})$ falls into the neighborhood $U(\theta, \frac{r}{\sqrt{n}})$, we have
\begin{equation}\label{normalized-likelihood}
p(x^{(n)};{\hat \theta}_n)
=e^{\{\frac{1}{2}[\sum_{i=1}^{n}\,S(x_i)-n\dot {A}(\theta)]^{T}[nI(\theta)]^{-1}[\sum_{i=1}^{n}\, S(x_i)-n\dot {A}(\theta)]+o(\frac{r}{\sqrt{n}})\}}p(x^{(n)};\theta)\, ,
\end{equation}
where ${\hat \theta}_n$ solves $\sum_{i=1}^{n}\,S(x_i)=n\dot {A}({\hat \theta}_n)$. Applying the Taylor expansion, we get
\[\sum_{i=1}^{n}\,S(x_i)-n\dot {A}(\theta)=n\dot {A}({\hat \theta}_n)-n\dot {A}(\theta)=[n\ddot {A}(\theta)]({\hat \theta}_n-\theta)+o(\frac{r}{\sqrt{n}})\, .
\]
Plugging it into (\ref{normalized-likelihood}) leads to
\begin{equation}\label{normalized-likelihood-by-density}
p(x^{(n)};{\hat \theta}_n)
=e^{\{\frac{1}{2}({\hat \theta}_n-\theta)^{T}[n\ddot {A}(\theta)]({\hat \theta}_n-\theta)+o(\frac{r}{\sqrt{n}})\}}p(x^{(n)};\theta)\, .
\end{equation}
If we consider i.i.d. random variables $Y_1, \cdots, Y_n$ sampled from the exponential distribution \eqref{exponential},
then the MLE $\hat \theta(Y^{(n)})$ is a random variable.
The summation of the quantity \eqref{normalized-likelihood-by-density} in the neighborhood $U(\theta, \frac{r}{\sqrt{n}})$
can be expressed as the following expectation of $\hat \theta(Y^{(n)})$.
\begin{equation}\label{expectation-cube}
E[e^{\{\frac{1}{2}({\hat \theta}_n-\theta)^{T}[n\ddot {A}(\theta)]({\hat \theta}_n-\theta)\}}{\bf 1}({\hat \theta}_n \in U(\theta, \frac{r}{\sqrt{n}}))]\, .
\end{equation}
Due to  the asymptotic normality of MLE $\hat \theta(Y^{(n)})$, namely,  ${\hat \theta}_n-\theta \mathop{\longrightarrow}\limits^{d} N(0,[nI(\theta)]^{-1})$, the  density of $\hat \theta(Y^{(n)})$ is approximated by
\[
\frac{|nI(\theta)|^{1/2}}{(2\pi)^{d/2}}\, e^{\{-\frac{1}{2}({\hat \theta}_n-\theta)^{T}[n\ddot {A}(\theta)]({\hat \theta}_n-\theta) \}} \, d{\hat \theta}_n\, .
\]
Applying this density to the expectation in \eqref{expectation-cube}, we find the two exponential terms cancel out, and obtain $\frac{|nI(\theta)|^{1/2}}{(2\pi)^{d/2}}$.
The sum of its
logarithm over all neighborhoods $U(\theta, \frac{r}{\sqrt{n}})$ leads to the remaining terms in \eqref{NML-codelength}.

\paragraph{Predictive coding}\label{inferential coding}
The optimality of the NML code is established in the minimax settings.
Yet its implementation requires going through the data two rounds, one for word counting of a dictionary, and one
more for encoding.
It is natural to ask whether there exists
a scheme that goes through the data only once and still can compress the data equally well.
It turns out that predictive coding is such a scheme for a given dictionary.
The idea of predictive coding is to sequentially make inferences about
the parameters in the probability function $p(x; \theta)$, which is then used to update
the code book.
That is, after obtaining observations $x_1, \cdots, x_i$, we
calculate the MLE
${\hat \theta}_{i}$, and in turn encode the next observation
according to the current estimated distribution.
Its code length is thus $L_{predictive}=(-\sum_{i=1}^{n}\, \log p(X_{i+1}|{\hat \theta}_{i}))$.
This procedure
(Rissanen \cite{Rissanen-86-ima,Rissanen-86-annals})  has
intimate connections with the prequential approach to
statistical inference as advocated by Dawid
\cite{Dawid-prequential,Dawid-in-spain}.
Predictive coding is intuitively optimal due to two important fundamental results.
First, the MLE ${\hat \theta}_{i}$ is asymptotically most accurate, since it gathers all the
information in $X_1, \cdots, X_i$ for the inference of the parametric model $p(x; \theta)$.
Second the code length $\log p(X_{i+1}|{\hat \theta}_{i})$ is optimal as dictated by the
Shannon source coding theorem. In the case of exponential families, Proposition 2.2 in \cite{my-2000-ieee} showed
$L_{predictive}$ can be expanded as follows.
\[
{L_{predictive}}=nH(\hat \theta_n)
+\frac{d}{2}\log \, n+{\tilde D}_n(\omega)
\, ,
\]
where the sequence of random variables $\{\tilde D_n(\omega)\}$
converges to an almost surely finite random variable $\tilde D(\omega)$.

Alternatively, we can use Bayesian estimates
in the predictive coding.
Starting from a prior distribution $w(\theta)$,
we encode $x_1$ by the marginal distribution
$q_1(x_1) = \int_{\Theta} p(x_1|\theta) w(\theta) d\theta$
resulted from $w(\cdot)$. The posterior
is given by $w_1(\theta)=p(x_1|\theta)
w(\theta)/\int_{\Theta} p(x_1|\theta) w(\theta) d\theta$.
We then use this posterior as the updated prior to encode the next word $x_2$.
Using induction, we can show that the marginal distribution to
encode the $k$-th word is
\[
q_k(x_k) = \frac{\int_{\Theta} [\prod_{i=1}^k\,
p(x_i|\theta)] w(\theta) d\theta}{\int_{\Theta} [\prod_{i=1}^{k-1}\,
p(x_i|\theta)] w(\theta) d\theta}\, .
\]
Meanwhile, the updated posterior, also the prior for the next round encoding, becomes
\[w_k(\theta)=\frac{[\prod_{i=1}^k\,
p(x_i|\theta)] w(\theta)}{\int_{\Theta} [\prod_{i=1}^{k}\,
p(x_i|\theta)] w(\theta) d\theta}\, .
\]
\begin{Prop} {\bf (Bayesian predictive code length)}
The total Bayesian  predictive code length for a string of $n$ words is
$${L_{mixture}}=-\sum_{k=1}^n\, \log q_k(x_k)=-
\log \int_{\Theta} [\prod_{i=1}^n\,
p(x_i|\theta)] w(\theta) d\theta\, .$$
\end{Prop}
Thus the Bayesian predictive code is nothing but the mixture code
referred to in the literature \cite{Clarke-Barron-jeffrey}.
\begin{Theo}{\bf (Expansion of Bayesian predictive code length)}\label{Bayesian-predictive-codelength-theorem}
If we fit a data sequence by an exponential family distribution, the mixture code length has
the expansion:
\begin{equation}\label{l-mixture}
{L_{mixture}}=nH({\hat \theta}_n)
+\frac{d}{2}\log \, \frac{n}{2\pi}
+\log \frac{|I({\hat \theta}_n)|^{1/2}}{w({\hat \theta}_n)}+o(1)\, ,
\end{equation}
where  $w(\theta)$ is any mixture of conjugate prior distributions.
\end{Theo}
The result is valid for general priors that can be approximated by a mixture of conjugate ones.
In the case of multinomial distributions, the conjugate prior is
Dirichlet distribution.
Any prior $w(\theta)$ continuous on the $d$-dimensional simplex in the $[0,1]^{(m+1)}$ cube can
be uniformly approximated by the Bernstein polynomials of $m$ variables, each term of
which is a  Dirichlet distribution \cite{powell-approximate,prolla_1988}.
It is noted that in the current setting, the source is not assumed to be
i.i.d. samples from an exponential family distribution as in Theorem 2.2 and Proposition 2.3
in \cite{my-2000-ieee}.

When $\int_{\Theta} |I(\theta)|^{1/2}\, d\theta$ is finite, we can take the Jeffreys prior,
$w(\theta)=\frac{|I(\theta)|^{1/2}}{\int |I(\theta)|^{1/2}\, d \theta}$
then
(\ref{l-mixture}) becomes (\ref{NML-codelength}). Putting together, we have shown
the optimal code length can be achieved by the Bayesian predictive coding scheme.

\paragraph{Redundancy}\label{Redundancy} Now we examine the empirical code length under Shannon's setting. That is,
 we evaluate the redundancy of
the code length assuming the source is from a hypothetical distribution.
\begin{Prop}
If we assume that a source follows an exponential family distribution, then
\begin{equation}\label{Difference between two entropies}
nH(\hat \theta_n)-nH(\theta)=-C_n(\omega)\, (\log\log \, n)+o(1)\, ,
\end{equation}
the sequence of nonnegative
random variables $\{C_n(\omega)\}$ have the property,
$\overline {\lim}_{n \rightarrow \infty}\,  C_n(\omega)\le d$,
for almost all path $\omega$'s.
If we further assume that $\int_{\Theta} |I(\theta)|^{1/2}\, d\theta<\infty$,
then
\begin{equation}\label{Redundancy of normalized likelihood code length}
R_{NML}=L_{NML}-NH(\theta)=\frac{d}{2}\log \, \frac{n}{2\pi}-C_n(\omega)\, (\log\log \, n)
+\log \int_{\Theta} |I(\theta)|^{1/2}\, d\theta+o(1)\, ,
\end{equation}
where $\{C_n(\omega)\}$ is the same as above.
\end{Prop}
The difference in the first part is
$(-\sum_{i=1}^{n}\, \log p(X_i|{\hat \theta}_{n}))-(-\sum_{i=1}^{n}\, \log p(X_i|\theta_0))$,
and the rest is true according to the proof of Proposition 2.2 in
\cite{my-2000-ieee}, Equation (18).
The NML code is a special case of the mixture code, whose
redundancy is given by Theorem 2.2 in \cite{my-2000-ieee}.
We note that $\overline {\lim}_{n \rightarrow \infty}\,  C_n(\omega)$ is bounded below by 1.
This proposition confirms that $nH(\hat \theta_n)$ is
an underestimate of the compression bound.
Although $\log\log \, n$ grows up slowly,
the term,  as shown by the
example in Table \ref{simulate-one}, gets large
as the model dimension $d$ increases.
\paragraph{Coding of discrete symbols and multinomial model}
For compressing strings of discrete symbols,
it is sufficient to consider the discrete distribution
specified by a probability vector,
$\theta=(p_1, p_2, \cdots, p_d, p_{d+1})$.
Its frequency function is
$P(X=k)=\prod_{k=1}^{d+1}\, p_k^{{\bf 1}(X=k)}$.
The Fisher information matrix
$I(p_1, \, \cdots\, p_d)$
can be shown to be
\[
-\left [ E\frac{\partial ^2{log P(X=k)}}{{\partial p_j}{\partial p_k}}\right ]_{j,k=1, \cdots, d}=
\left ( \begin{array}{cccc}
\frac{1}{p_1}+\frac{1}{p_{d+1}}&\frac{1}{p_{d+1}}&\cdots&\frac{1}{p_{d+1}}\\
\frac{1}{p_{d+1}}&\frac{1}{p_2}+\frac{1}{p_{d+1}}&\cdots&\frac{1}{p_{d+1}}\\
\vdots&\vdots&\ddots&\vdots\\
\frac{1}{p_{d+1}}&\frac{1}{p_{d+1}}&\cdots&\frac{1}{p_d}+\frac{1}{p_{d+1}}\\
\end{array}\right)\, .
\]
Thus $|I(p_1, \, \cdots\, p_d)|=1/\prod_{k=1}^{d+1} p_k$.

Suppose $X_1, \cdots, X_n$ are i.i.d. random variables obeying the above
discrete distribution. Then $S=\sum_{i=1}^{n}\, X_i$  follows a multinomial distribution
$Multi (n; p_1, p_2, \cdots, p_d, p_{d+1})$. Its conjugate prior distribution is  the Dirichlet distribution
$(\alpha_1, \alpha_2, \cdots, \alpha_{d+1})$, whose density function is
$$
\frac{\Gamma(\sum_{k=1}^{d+1} \alpha_k)}{\prod_{k=1}^{d+1}\Gamma(\alpha_k)}
\prod_{k=1}^{d+1}\, p_k^{\alpha_k-1}\, ,
$$
where $\Gamma(t)=\int_{u=0}^{+\infty} u^{-t} e^{-u} du$.
Since the Jeffreys prior is proportional to $|I(p_1, \, \cdots\, p_d)|^{1/2}$.
In this case, it equals Dirichlet(1/2,1/2, ..., 1/2), whose
density is $\frac{\Gamma((d+1)/2)}{\Gamma(1/2)^{d+1}}\prod_{k=1}^{d+1}\, p_k^{-1/2}$.
The Jeffreys prior was also used by
Krichevsky \cite{Krichevskii_1968} to derive optimal universal codes.

It is noticed that $\Gamma(1/2)={\sqrt \pi}$.
Plug it into Equation \eqref{l-mixture},
we have the following specific form of the NML code length for the multinomial distribution.
Remember that the distribution or word frequencies are specific for a given dictionary ${\Phi}$, and we thus
term it as $L_{NML@{\Phi}}$. If we change the dictionary, the code length changes accordingly.
\begin{Prop}{\bf (Optimal code lengths for a multinomial distribution)}
\begin{equation}\label{empirical-multinomial}
L_{NML@{\Phi}}=nH(\hat \theta_n)
+\frac{d}{2}\log \, n
-\frac{d}{2}
-\log \Gamma(\frac{d+1}{2})+\frac{1}{2}\log \pi \, ,
\end{equation}
where $nH(\hat \theta_n)=
-\sum_{k=1}^{d} {\hat p}_k \log {\hat p}_k$, ${\hat p}_k=n_k/n$---
the frequency of the k-th word appearing in the string.
\end{Prop}

\section{Compression of random sequences and DNA sequences}
\paragraph {Lossless compression bound and description length}
Given a dictionary of words, we parse a string into words followed by
counting their frequencies ${\hat p}_k=n_k/n$,
the total number of words $n$, and the number of distinct words $d$. Plugging them into
expression (\ref{empirical-multinomial}), we obtain the lossless compression bound for this
dictionary or parsing. If a different parsing is tried, the three quantities: word frequencies, number of words,
dictionary size (number of distinct words) would change, and the resulting bound would change accordingly.

In the general situation where the data are not necessarily discrete symbols,
we replace the code length with description length \eqref{l-mixture} as termed by Rissanen.

Since each parsing corresponds to a probabilistic model, the code length is model-dependent.
The comparison of two or more coding schemes is exactly the selection of
models, with the expression (\ref{empirical-multinomial}) as the
target function.
\paragraph {Rissanen's principle of minimum description length and model selection}\label{Rissanen's work}
Rissanen, in his work of
\cite{Rissanen-86-annals,Rissanen-book,Rissanen-96}, etc.
proposed the principle of minimum description length (MDL) as
a more general modeling rule than that
of maximum likelihood, which
was recommended, analyzed, and popularized by R. A. Fisher.
From the information-theoretic point of view,
when we encode data from a source by prefix coding,
the optimal code is the one that achieves the minimum
description length.
Because of the equivalence  between
a prefix code length and the negative
logarithm  of the corresponding
 probability distribution, via Kraft's inequality,
this in turn gives us a modeling principle,
namely, the MDL principle: choose the model or prefix coding algorithm that
gives the minimal description of data, see Hansen and Yu \cite{Hansen-Yu-98}
for a  review on this topic. We also refer readers to \cite{Barron-R-Yu-98,Grunwald2007}
for a more complete account of MDL.

MDL is a mathematical formulation of the general principle known as Occam's razor:
choose the simplest explanation consistent with
the observed data \cite{Cover-book}.
We make one remark about the significance of MDL.
On the one hand, Shannon's work establishes the connection
between optimal coding and probabilistic models.
On the other hand, Kolmogorov's algorithmic theory
says that the complexity, or the absolute optimal coding,
cannot be proved by any Turing machines.
MDL offers a practical principle: it allows us to make
choices among possible models and coding algorithms without the desire to
prove optimality. As more and more candidates of models are evaluated over time,
human's understanding progresses.

\paragraph{Compression bounds of random sequences}
A random sequence is non-compressible by any model-based or algorithmic prefix
coding as indicated by the complexity results \cite{Li-Vat-book, Li-Vat-2000-ieee}.
Thus a legitimate compression bound of a random sequence should be no less than 1 up to
certain variations.
Conversely,  if the compression rates of a sequence using $L_{NML@{\Phi}}$ as the compression bound
is no less than 1 under all dictionaries ${\Phi}$, namely,
\[
\min_{{\Phi}: dictionaries}\, \frac{L_{NML@{\Phi}}}{L_{RAW}}=1+\frac{L_{NML@{\Phi}}-L_{RAW}}{L_{RAW}}\ge 1\, ,
\]
where $L_{RAW}$ is the data length of the raw sequence in terms of bits, then the sequence is random.
If we assume the source is from a uniform distribution, $L_{RAW}=nH$, and
the difference $L_{NML@{\Phi}}-L_{RAW}$ is essentially the
redundancy of $L_{NML@{\Phi}}$, which can be calculated by \eqref{empirical-multinomial}.
Although it is challenging to test all dictionaries, we can try some, particularly those
suggested by the domain experience.

\paragraph{A simulation study: compression bounds of pseudo-random sequences}
Simulations were carried out to test the theoretical bounds.
First, a pseudo-random binary string of size 3000 was simulated in R according to Bernoulli trials with a probability of 0.5.
In Table \ref{simulate-one}, the first column shows the word length used for parsing the data; The second column
shows the word number; the third column shows the number of distinct words.
We group the terms in \eqref{empirical-multinomial} into three parts: the term involving $n$, the term involving $\log n$, and others.
The bounds by $nH(\hat \theta_n)$, $nH(\hat \theta_n)+\frac{d}{2}\log \, n$ and $L_{NML}$ in \eqref{empirical-multinomial}
are respectively shown in the next three columns.
When the word length increases, $d$ increases, and the bounds by $nH(\hat \theta_n)$ show a decreasing trend.
A bound smaller than 1 indicates the sequence
can be compressed,  contradicting the assertion that random sequences cannot be so.
When the word length is 8, the dictionary size is 375, and the bound by $nH(\hat \theta_n)$ is only 0.929.
The incompressibility nature of random sequences falsified
$nH(\hat \theta_n)$ as a legitimate compression bound.
If the $\log n$ term is included, the bounds are always larger than 1.
The bounds by $L_{NML}$ \eqref{empirical-multinomial} are tighter while remain larger than 1 except
the case at the bottom row, where the number of distinct words approaches the total number of words.
Since $L_{NML}$ is an achievable bound, $nH(\hat \theta_n)+\frac{d}{2}\log \, n$ is an overestimate.
\begin{table}[tbp]
\caption{The data compression rates of a binary string of size 3000 under different parsing models. The data were simulated in R
according to Bernoulli trials with probability 0.5.} \label{simulate-one}
  \begin{center}
    \begin{tabular}{|r|r|r|r|r|r|}
    \hline
        word & word  & dictionary  & {$nH(\hat \theta_n)$}  & $nH(\hat \theta_n)$ & $L_{NML}$ \eqref{empirical-multinomial} \\
      length & number & size &   & $+\frac{d}{2}\log \, n$  &
         \\ \hline
        1 & 3000 & 2 & 0.999918 & 1.001843 & 1.001952 \\ \hline
        2 & 1500 & 4 & 0.998591 & 1.003866 & 1.003641 \\ \hline
        3 & 1000 & 8 & 0.998961 & 1.010587 & 1.008834 \\ \hline
        4 & 750 & 16 & 0.995422 & 1.019299 & 1.012974 \\ \hline
        5 & 600 & 32 & 0.989603 & 1.037285 & 1.018977 \\ \hline
        6 & 500 & 64 & 0.981597 & 1.075738 & 1.027959 \\ \hline
        7 & 428 & 124 & 0.964885 & 1.144324 & 1.031261 \\ \hline
        8 & 375 & 196 & 0.928930 & 1.206829 & 1.006312 \\ \hline
        9 & 333 & 252 & 0.872958 & 1.223846 & 0.950283\\ \hline
    \end{tabular}
    \end{center}
\end{table}

\paragraph{Knowledge discovery by data compression}
On the other hand, if we can compress a sequence by a certain prefix coding scheme, then
this sequence is not random. In the meantime,  this coding scheme
presents a clue to understanding the information structure hidden in the sequence.
Data compression is one general learning mechanism, among others, to discover knowledge
from nature and other sources.

Ryabko,  Astola and Gammerman \cite{Ryabko_Astola_Gammerman_2006}
applied the idea of Kolmogorov complexity to the statistical testing
of some typical hypotheses.
This approach was used to analyze DNA sequences in \cite{Usotskaya_Ryabko_2009}.

\paragraph{DNA sequences of proteins}
The information carried by the DNA double helix
is two long complementary strings of the letters A, G, C, and T.
It is interesting to see if we can compress DNA sequences
at all. Next, we carried out the lossless compression experiments on a couple of protein-encoding DNA sequences.
\paragraph {Rediscovery of the codon structure}\label{genes}
In Table \ref{first-one} we show the result by applying  the NML code length $L_{NML}$ in \eqref{empirical-multinomial}
to an {\it E. Coli} protein gene sequence labeled by {\it b0059} \cite{Ecoli-genome}, which has 2907 nucleotides.
Each row corresponds to one model used for coding.
All the models being tested are listed in the first column. In the first model,
we encode the DNA nucleotides one by one and name it Model 1. In the second or third model,
we parse the DNA sequence by pairs and then encode the resulting bi-nucleotide sequence according to
their frequencies.
Different starting positions lead to two different phases, and
we denote them by 2.0 and 2.1 respectively.
Other models are understood in the same fashion. Note
that all these models are generated by fix-length parsing.
The last model ``a.a.'' means we first translate DNA triplets into
amino acids and then encode the resulting amino acid sequence.
The second column shows the total number of words in each parsed sequence.
The third column shows the number of different words in each parsed sequence
or the dictionary size.
The fourth column is the empirical entropy estimated from observed frequencies.
The next column is the first term in expression \eqref{empirical-multinomial}, which is the
product of the second and fourth columns. Then we calculate the rest terms in
\eqref{empirical-multinomial}. The total bits are then calculated and the compression
rates are the ratios $L_{NML}/(2907*2)$. The last column shows the compression rates under different models.

All the compression rates are around 1 except that obtained
from Model 3.0, which
represents the correct codon pattern and correct phase.
Thus the comparison of compression bounds rediscovers  the codon structure of this protein-encoding DNA sequence and
the phase of the open reading frame.
It is somewhat surprising that the optimal code length $L_{NML}$ enables  us to
 mathematically identify the
triplet coding system using only the sequence of one gene.
Historically, the system was
discovered by Francis Crick and his colleagues in the early 1960s
using frame-shift mutations of bacteria-phage T4.

Next, we have a closer look at the results.
The compression rate of the four-nucleotide word coding is closest to 1, and
thus it behaves more like "random". For example,
it is 0.9947 for Model 4.2. The first term of empirical entropy contributes 5431 bits,
while the rest terms contribute 346 bits. If we use  $\frac{d}{2}\log \, n$ instead,
the rest term is
$0.5*(219-1)*\log(726)\approx 1036$ bits, and the compression rate becomes 1.11, which is less tight.
If the Ziv-Lempel algorithm is applied to the {\it b0059} sequence,
635 words are generated along the way.
Each word needs $log(635)$ bits for keeping the address of its prefix, and 2 bits for the last nucleotide.
In total, it needs $635*log(635)=5912$ bits for storing addresses, which correspond
to the first term in \eqref{empirical-multinomial}, and $635*2=1270$ bits for storing the words' last
letter, which correspond to the rest terms in \eqref{empirical-multinomial}.
The compression rate of Ziv-Lempel coding is 1.24.

{\footnotesize
\begin{table}[btp]
\caption{The data compression rates of  {\it E. Coli} ORF {\it b0059}
calculated by \eqref{empirical-multinomial} under different parsing models.} \label{first-one}
  \begin{center}
    \begin{tabular}{||r|r|r|r|r|r|r|r||}
\hline
model & \# word & dictionary & empirical & 1-st & rest  & $L_{NML@{\Phi}}$  & compression \\
 & $n$ & size $d$ & entropy & term & terms & total bits &  rate\\

\hline
  1    &    2907  &    4  &  1.9924 &  5792.00  &   16.58  &   5808.58 &  0.9991\\
\hline
    2.0   &    1453  &    16 &    3.9570 &   5749.49  &    59.81 &    5809.31&   0.9995\\
\hline
  2.1    &    1453   &    16   &   3.9425  &   5728.44  &    59.81  &  5788.25 & 0.9959\\
\hline
{\bf   3.0}  &   {\bf 969} &  {\bf 58}  & {\bf 5.2842} & {\bf 5120.39}    &   {\bf 157.11} &  {\bf 5277.51} & {\bf 0.9077}\\
\hline
  3.1   &    968  &   63 &   5.5905 &  5411.63 &   167.13  &  5578.76 & 0.9605\\
\hline
  3.2     &  968   &  64  &  5.6706 &  5489.10  &  169.11 &   5658.21 & 0.9742\\
\hline
  4.0  &     726  &   218   &  7.4507  &  5409.24   &  345.07  &   5754.31 & 0.9908\\
\hline
  4.1   &    726  &  217  &  7.4337 &  5396.87 &   344.20 &   5741.07 & 0.9885\\
\hline
  4.2    &   726 &   219 &   7.4814 &  5431.49 &   345.940 &   5777.43 & 0.9947\\
\hline
  4.3     &  726 &   221  &  7.4678 &  5421.64 &   347.67 &   5769.31 & 0.9933\\
\hline
a. a.   &  969 &    21  &  4.1056 &  3978.31  &   69.92  &  4048.22 & 0.6963\\
\hline
   \end{tabular}
  \end{center}
\end{table}
}

\paragraph {Redundant information in protein gene sequences}\label{redundancy}
It is known that the $4^3=64$ triplets correspond to only 20 amino acids plus stop codons. Thus redundancy does
exist in protein gene sequences. Most of the redundancy lies in the third position
of a codon. For example, GGA, GGC, GGT, and GGG all correspond to glycine.
According to Table \ref{first-one}, there are 4048.22 bits of information in the amino acid
sequence while there are 5277.51 bits of information in  Model 3.0.
Thus the redundancy in this sequence is estimated to be (5277.51-4048.22)/4048.22=0.30.

\paragraph{Randomization}\label{Randomization}
To evaluate the accuracy or significance of the compression rates of a DNA sequence, we need
a  reference distribution for comparison.
A typical method is
to consider the randomness obtained by permutations. That is, give a DNA sequence, we permute
the nucleotide bases and re-calculate the compression rates. If we repeat this permutation procedure,
then a reference distribution is generated.

In Table \ref{second-one}, we consider the
compression rates for {\it E. Coli} ORF {\it b0060}, which has 2352 nucleotides.
First, the optimal compression rate of 0.958 is achieved at model 3.0.
Second, we further carried out the calculations
for permuted sequences. The averages, standard deviations, and 1$\%$ lower  quantiles of compression rates under different
models are shown in
Table \ref{second-one} as well.
Except for Model 1, all the compression rates, in terms of either averages or lower quantiles are significantly above 1,
Third, the results by the single term $nH(\hat \theta_n)$ are about
0.996, 0.994, 0.986, and 0.952 respectively for one-, two-, three-, and four-nucleotide models.
The $99\%$ quantiles of $nH(\hat \theta_n)$ for the four-nucleotide models are no larger than 0.961.
Thus  the
difference between $nH(\hat \theta_n)$ and $nH(\theta)$ as shown in \eqref{Difference between two entropies}
increases as the dictionary size goes up.
Fourth, the results of $nH(\hat \theta_n)+\frac{d}{2}\log \, n$
show extra bits compared to those of $L_{NML}$, and the compression ratio go from
1.02 to 1.17, suggesting the rest terms in \eqref{empirical-multinomial} are not
negligible.

{\footnotesize
\begin{table}[tbp]
\caption{The data compression rates of  {\it E. Coli} ORF {\it b0060} and statistics from  permutations.
 The protein gene sequence has 2352 nucleotide bases.
} \label{second-one}
  \begin{center}
  {\scriptsize
    \begin{tabular}{||r|r|r|r|r|r|r|r|r|r|r|r||}
\hline
Model& & 1.0 & 2.0 & 2.1 & 3.0 & 3.1 & 3.2 & 4.0 & 4.1 & 4.2 & 4.3\\
\hline
Original & $L_{NML}$ \eqref{empirical-multinomial} &
0.999 &   1.000 &   1.001 &   0.958 &  0.980 &   0.989 &   1.001 &   1.002 &   0.997 &   0.999
\\
\hline
\multirow{3}{*}{$L_{NML}$ \eqref{empirical-multinomial}} & average&
0.999 &           1.006&             1.006&             1.020&             1.020&
1.020&           1.020&             1.020 &             1.020&             1.020\\
&SD ($\times 10^{-3}$) &
0.00 &        0.74     &       0.76    &        1.69   &         1.77  &
1.71 &          4.22     &        4.33    &        4.15   &          3.92 \\
&$1\%$-quantile  &
 0.999     &         1.004   &            1.004  & 1.016  &             1.016     &          1.016 &
 1.011     &        1.010      &         1.011 &  1.011
\\
\hline

\multirow{ 3}{*}{$nH(\hat \theta_n)$} &average&
0.996   &       0.994   &       0.994        &     0.986   &       0.987&
0.986   &        0.952   &          0.952     &        0.952   &         0.952 \\
&SD ($\times 10^{-3}$) &
0.00 &           0.74     &      0.76   &        1.69   &        1.77 &
1.71    &       3.69    &       3.79    &       3.63   &        3.42\\
& $99\%$-quantile  &

0.996     &        0.995  &            0.995    &          0.990   &           0.990&

0.990    &      0.961   &           0.961    &          0.960      &        0.960 \\

\hline

\multirow{2}{*}{$nH(\hat \theta_n)$}
&average&
0.999     &         1.010  &             1.010  &             1.051  &             1.051 &
1.051         &      1.174        &       1.174   &            1.174 &              1.174\\
&SD ($\times 10^{-3}$) &
0.00   &    0.74  &        0.76  &       1.70   &       1.78 &
1.71    &     7.49   &      7.66   &       7.37  &        7.01 \\
$+\frac{d}{2}\log \, n$ &$1\%$-quantile  &
0.999     &      1.008    &           1.008      &         1.046   &         1.047&
1.047   &          1.157   &            1.156    &          1.157    &         1.157\\
\hline

   \end{tabular}
   }
  \end{center}
\end{table}
}

It is noted Models 3.1 and 3.2 are obtained by phase-shifting from the
correct Model 3.0. Other models are obtained by incorrect parsing.
These models can serve as references for  Model 3.0.
The incorrect parsing and phase-shifting have a flavor of the linear congruential pseudo-random number generator,
and play the role of randomization.

\section{Discussion}
Putting together the analytical results and numerical examples, we show
the compression bound of a data sequence using
an exponential family is the code length derived from the NML distribution \eqref{NML-codelength}.
The empirical bound can be implemented
by the Bayesian predictive coding for any given dictionary or model.
Different models are then compared by their empirical compression bounds.

The examples of DNA sequences indicate that
the compression rates by any dictionaries are indeed larger than
1 for random sequences, in consistency with
the assertions by the Kolmogorov complexity theory.
Conversely, if
significant compression is achieved by a specific model, certain knowledge is gained.
The codon structure is such an instance.

Unlike the algorithmic complexity that contains a constant, the results based on
probability distributions give the exact bits of code lengths.
All three terms in \eqref{NML-codelength} are important for
the compression bound.
Using only the  first term $nH(\hat \theta_n)$
can lead to bounds of random sequences smaller than 1.
The gap gets larger as the dictionary size increases as seen
from Table \ref{simulate-one} and \ref{second-one}.
The bound by adding the second term $\frac{d}{2}\log \,n$
had been proposed by the two-part coding or the Kolmogorov complexity.
It is equivalent to BIC widely used in model selection.
However, it overestimates the influence of the dictionary size, as shown by the
examples of simulations and DNA sequences.
The inclusion of the Fisher information in the third term gives a tighter bound.
The terms other than $nH(\hat \theta_n)$ get larger as the dictionary size increases
in Table \ref{simulate-one} and \ref{second-one}.
The observation that the compression bounds from all terms in \eqref{NML-codelength} kept slightly
above 1 for all tested libraries meets our expectation on the incompressibility of random sequences.

Although the empirical compression bound is obtained under the i.i.d. model,
the word length can be set rather large to describe the local dependence
between symbols. Indeed, as shown in the examples of DNA sequences,
the empirical entropy term in \eqref{NML-codelength}
could get smaller, for either
the original sequences or the permuted ones. Meanwhile, the second term
could get larger. For a specific sequence, a better dictionary
is selected by trading off the entropy part and model complexity part.

Rissanen \cite{Rissanen-96} obtained an expansion of the
NML code length, in which
the first term is the log-likelihood of data with the parameters plugged in by the MLE.
In this article, we show it is exactly the empirical entropy if the parametric model takes
any exponential family.
According to this formula, the NML code length
is an empirical version or a direct extension of Shannon's source coding theorem.
Furthermore, the asymptotics in \cite{Rissanen-96}
requires five assumptions, which are hard to examine.
 Suzuki and Yamanishi proposed a Fourier approach to calculate
 NML code length \cite{8437862} for continuous random variables with certain assumptions.
Instead, we show \eqref{NML-codelength} is valid
for exponential families, as long as $\int_{\Theta} |I(\theta)|^{1/2}\, d\theta<\infty$, without any other assumptions.
If the Jeffreys prior is improper in the interior of the full parameter space, we can restrict the parameter to a compact subset.
The exponential families include not only distributions of discrete symbols such as multinomials but also
continuous signals such as from the normal distribution.

The mathematics underlying the expansion of NML is
the structure of local asymptotic
normality proposed by LeCam \cite{LeCam-Yang-book}.
LAN has been used to
show the optimality of certain statistical estimates.
This article connects LAN to
compression bound.
We have shown as long as LAN is valid, the similar expansion of \eqref{empirical-likelihood-logsum}
can be obtained.

\section {Appendix}\label{Appendix}
This section contains the proofs of the results in Sections 2 and 3.
The following basic facts about the exponential family
(\ref{exponential}) are needed, see \cite{10.5555/41464}.

\begin{enumerate}
\item
$E(S(X))=\dot {A}(\theta)$, and $Var(S(X))=\ddot {A}(\theta)$.
\item $\dot {A}(\cdot)$ is one to one on the natural parameter space.
\item The MLE $\hat{\theta}_n$ based on
$(X_1, \cdots, X_n)$ is given by
%\begin{equation}\label{mle-formula}
$\hat{\theta}_n=\dot {A}^{-1}(\bar{S}_n)$,
%\end{equation}
where $\bar{S}_n=\frac{1}{n}\sum_{i=1}^{n}\, S(X_i)$.
\item The Fisher information matrix $I(\theta)=\ddot {A}(\theta)$.
\end{enumerate}

\n {\bf Proof of Theorem \ref{main-theorem}}.
In the canonical exponential family,  the natural parameter space is open and convex.
Since $\int_{\Theta} |I(\theta)|^{1/2}\, d\theta<\infty$,
we can find a series of
bounded set $\{\Theta^{(k)}, k=1,2,\cdots\}$ such that
$\log[\int_{\Theta} |I(\theta)|^{1/2}\, d\theta]-\log[\int_{\Theta^{(k)}} |I(\theta)|^{1/2}\, d\theta]=\epsilon_k$.
where $\epsilon_k\rightarrow 0$.
Furthermore, we can select each bounded set $\Theta^{(k)}$ so that it can be partitioned into disjoint
cubes, each of which is denoted by $U(\theta_j^{(k)}, \frac{r}{\sqrt{n}})$ with $\theta_j^{(k)}$  as its center and $\frac{r}{\sqrt{n}}$ as its side length.
Namely,
$\Theta^{(k)}=\bigcup_j U(\theta_j^{(k)}, \frac{r}{\sqrt{n}})$, and $U(\theta_{j_1}^{(k)}, \frac{r}{\sqrt{n}})\bigcap U(\theta_{j_2}^{(k)}, \frac{r}{\sqrt{n}})=\emptyset$
for $j_1\neq j_2$.

The normalizing constant in equation (\ref{entropy-logsum}) can be summed (integration in the case of continuous variables) by the sufficient statistic
$\sum_{i=1}^{n} S(x_i)$, and in turn by the MLE ${\hat \theta}_n$

\begin{equation}\label{double-sum}
{\sum_{\{x^{(n)}, \, \hat \theta\in \Theta^{(k)}\}}\,p(x^{(n)};{\hat \theta}_n)}=\sum_{U(\theta_j^{(k)}, \frac{r}{\sqrt{n}})}
\sum_{\{x^{(n)}, \, \hat \theta \in U(\theta_j^{(k)}, \frac{r}{\sqrt{n}})\}}  \, p(x^{(n)};{\hat \theta}_n)\, ,
\end{equation}

\begin{equation}\label{maximized-likelihood}
p(x^{(n)};{\hat \theta}_n)=e^{\{{\hat \theta}_n^T \sum_{i=1}^{n}\,S(x_i)-nA({\hat \theta}_n)\}} \mu^n (dx^n)
=e^{\{n{\hat \theta}_n^T \bar{S}_n-nA({\hat \theta}_n)\}} \mu^n (dx^n)\, .
\end{equation}
Now  expand $n[\theta \bar{S}_n-A(\theta)]$ around ${\hat \theta}_n$ within the neighborhood $U(\theta_j^{(k)})$.
$$
n[\theta^T \bar{S}_n-A(\theta)]=n[{\hat \theta}_n^T \bar{S}_n-A({\hat \theta}_n)]+(\theta-{\hat \theta}_n)^T[n\bar{S}_n-n\dot A ({\hat \theta}_n)]
-\frac{1}{2}(\theta-{\hat \theta}_n)^T [n\ddot A ({\hat \theta}_n)] (\theta-{\hat \theta}_n)+M_1n||\theta-{\hat \theta}_n ||^3\, .
$$
Since the MLE ${\hat \theta}_n=\dot A ^{-1}(\bar{S}_n)$, the second term is zero. Furthermore, we expand
$\ddot A (\hat \theta_n)$ around $\ddot A (\theta)$, and rearrange the terms in the equation, then we have

$$
n[{\hat \theta}_n^T \bar{S}_n-A({\hat \theta}_n)]=n[\theta^T \bar{S}_n-A(\theta)]+\frac{1}{2}({\hat \theta}_n-\theta)^T [n\ddot A (\theta)] ({\hat \theta}_n-\theta)+
M_2 n||\theta-{\hat \theta}_n ||^3\, ,
$$
where the constant $M_2$ involves the
third order derivatives of $A(\theta)$, which is continuous in the canonical exponential family and thus
bounded in the bounded set $\Theta^{(k)}$. In other words, $M_2$ is bounded uniformly across all
$\{U(\theta_j^{(k)}, \frac{r}{\sqrt{n}})\}$. Similar bounded constants will be used repeatedly hereafter.
Then equation (\ref{maximized-likelihood}) becomes
$$
p(x^{(n)};{\hat \theta}_n)
=e^{\frac{1}{2}(\theta-{\hat \theta}_n)^T [n\ddot A (\theta)] (\theta-{\hat \theta}_n)+M_2 n||\theta-{\hat \theta}_n ||^3} e^{n[\theta \bar{S}_n-A(\theta)]} \mu^n (dx^n)\, ,
$$
Notice that the exponential form $e^{n[\theta \bar{S}_n-A(\theta)]}$ is the density of
$n\bar{S}$. If we consider i.i.d. random variables $Y_1, \cdots, Y_n$ sampled from the exponential distribution \eqref{exponential},
the MLE $\hat \theta(Y^{(n)})$ is a random variable.
Take $\theta=\theta_j^{(k)}$, then the sum of the above quantity over the neighborhood $U(\theta_j^{(k)}, \frac{r}{\sqrt{n}})$
is nothing but the expectation of $\hat \theta(Y^{(n)})=\dot A ^{-1}(\bar{S}_n)$ with respect to the distribution of
$n\bar{S}_n$, evaluated at the parameter $\theta_j^{(k)}$.

$$\sum_{\{x^{(n)}, \, {\hat \theta}_n \in U(\theta_j^{(k)}, \frac{r}{\sqrt{n}})\}}  \, p(x^{(n)};{\hat \theta}_n)
=E_{\theta_j^{(k)}}[{\bf 1}[{\hat \theta}_n \in U(\theta_j^{(k)}, \frac{r}{\sqrt{n}})]e^{\frac{1}{2}({\hat \theta}_n-\theta_j^{(k)})^T [n\ddot A (\theta_j^{(k)})] ({\hat \theta}_n-\theta_j^{(k)})+M_2 n||{\hat \theta}_n-\theta_j^{(k)} ||^3}]\, .
$$

Let $\xi_n=\sqrt{n} ({\hat \theta}_n-\theta_j^{(k)})$. Now ${\hat \theta}_n \in U(\theta_j^{(k)}, \frac{r}{\sqrt{n}}$) if and only if $\xi_n \in U(0, r)$,
where $U(0, r)$ is the d-dimensional cube centered at zero with the side length $r$. Next expand $e^{M_2 n||{\hat \theta}_n-\theta_j^{(k)} ||^3}$
in the neighborhood, the above becomes

$$\sum_{\{{\hat \theta}_n \in U(\theta_j^{(k)}, \frac{r}{\sqrt{n}})\}}  \, p(x^{(n)};{\hat \theta}_n)
=E[{\bf 1}[\xi_n \in U(0, r)]
[e^{\frac{1}{2}\xi_n^T I(\theta_j^{(k)}) \xi_n}(1+M_3 n^{-\frac{1}{2}})]\, .
$$
According to the central limit theorem,
$n^{-\frac{d}{2}}[\sum_{i=1}^{n}\,S(Y_i)- \dot A(\theta_j^{(k)})]\mathop{\longrightarrow}\limits^{d} N(0,\ddot A(\theta_j^{(k)}))$.
Moreover, the approximation error has the Berry-Esseen bound $O(n^{-\frac{1}{2}})$, where the constant is determined by the bound
on $A(\theta)$'s third-order derivatives. Similarly, we have
the asymptotic normality of MLE, $\xi_n(Y^{(n)})\mathop{\longrightarrow}\limits^{d} N(0,I(\theta_j^{(k)})^{-1})$, where the
Berry-Esseen bound is valid for the convergence,
see \cite{10.1214/16-EJS1133}. Therefore,
the expectation converges as follows.
\begin{eqnarray*}
&& E
\{ {\bf 1}[\xi_n \in U(0, r)]
e^{\frac{1}{2}\xi_n^T I(\theta_j^{(k)}) \xi_n} (1+M_3 n^{-\frac{1}{2}})\}\\
&=&
\int\, {\bf 1}[\xi_n \in U(0, r)]
[e^{\frac{1}{2}\xi_n^T I(\theta_j^{(k)}) \xi_n}(1+M_3 n^{-\frac{1}{2}})]
\frac{|I(\theta_j^{(k)})|^{1/2}}{(2\pi)^{d/2}}\,e^{-\frac{1}{2}\xi_n^T I(\theta_j^{(k)}) \xi_n} \, d\xi_n+M_4 n^{-\frac{1}{2}}
\nonumber \\
&=&{(2\pi)^{-\frac{d}{2}}} |I(\theta_j^{(k)})|^{\frac{1}{2}}\int\, {\bf 1}[\xi_n \in U(0, r)]
(1+M_3 n^{-\frac{1}{2}})] d\xi_n+ M_4 n^{-\frac{1}{2}}
\nonumber \\
&=&{(2\pi)^{-\frac{d}{2}}} |I(\theta_j^{(k)})|^{\frac{1}{2}} r^d (1+M_3n^{-\frac{1}{2}})+ M_4 n^{-\frac{1}{2}}
\\
&=&n^{\frac{d}{2}}{(2\pi)^{-\frac{d}{2}}} |I(\theta_j^{(k)})|^{1/2} (r^d n^{-\frac{d}{2}}) (1+M'_3 n^{-\frac{1}{2}})\, .
\end{eqnarray*}

Plug this into the sum (\ref{double-sum}), we obtain
\begin{eqnarray*}
&&{\sum_{\{x^{(n)}, \, {\hat \theta}_n\in \Theta^{(k)}\}}\,p(x^{(n)};{\hat \theta}_n)}=n^{\frac{d}{2}}{(2\pi)^{-\frac{d}{2}}} [\sum_{U(\theta_j^{(k)}, \frac{r}{\sqrt{n}})}\,  |I(\theta_j^{(k)})|^{1/2} (r^d n^{-\frac{d}{2}})] (1+M'_3 n^{-\frac{1}{2}})
\\
&\longrightarrow& n^{\frac{d}{2}}{(2\pi)^{-\frac{d}{2}}} [\int_{\Theta^{(k)}} |I(\theta)|^{1/2}\, d\theta] (1+M'_3 n^{-\frac{1}{2}})]\
%\mathop{\longrightarrow}\limits^{k\rightarrow \infty} n^{\frac{d}{2}}{(2\pi)^{-\frac{d}{2}}} \int_{\Theta} I(\theta)|^{1/2}\, d\theta\\
\end{eqnarray*}

\begin{eqnarray*}
&&\log[{\sum_{\{x^{(n)}, \, {\hat \theta}_n\in \Theta^{(k)}\}}\,p(x^{(n)};{\hat \theta}_n)}]=
\frac{d}{2}\log \, \frac{n}{2\pi}+\log \int_{\Theta^{(k)}} |I(\theta)|^{1/2}\, d\theta+ M''_3 n^{-\frac{1}{2}}\\
&=& \frac{d}{2}\log \, \frac{n}{2\pi}+\log \int_{\Theta} |I(\theta)|^{1/2}\, d\theta+[\log \int_{\Theta^{(k)}} |I(\theta)|^{1/2}\, d\theta-
\log \int_{\Theta} |I(\theta)|^{1/2}\, d\theta]+
+ M''_3 n^{-\frac{1}{2}}\\
&=& \frac{d}{2}\log \, \frac{n}{2\pi}+\log \int_{\Theta} |I(\theta)|^{1/2}\, d\theta-\epsilon_k+ M''_3 n^{-\frac{1}{2}}\, .
\end{eqnarray*}
Note that the bound $M''_3$ of the last term relies solely on $\Theta^{(k)}$. For a given $k$, we select $n$ such that the last term is
sufficiently small. This completes the proof.

\vspace{0.2cm}

\n {\bf Proof of Theorem \ref{Bayesian-predictive-codelength-theorem}}
First we consider the conjugate prior of (\ref{exponential}), which takes the form
\begin{equation}\label{conjugate}
u(\theta)=\exp\{\alpha'\,\theta\,-\beta\,A(\theta)-B(\alpha,\beta)\}\, ,
\end{equation}
where $\alpha$ is a vector in ${\bf R}^d$, $\beta$ is a scalar, and
$B(\alpha,\beta)=\log \, \int_{\Theta} \exp\{\alpha'\,\theta\,-
\beta\,A(\theta)\}\,d\theta$.
Then the marginal density is
\begin{equation}\label{conjugate-marginal}
m(x^{(n)})=\exp\{B(\sum_{i=1}^nT(x_i)+\alpha,n+\beta)-
B(\alpha,\beta)\}\, ,
\end{equation}
according to the definition of $B(\cdot, \cdot)$. Therefore
\[
{L_{mixture}}=B(\alpha,\beta)-B(\sum_{i=1}^nS(x_i)+\alpha,n+\beta)=B(\alpha,\beta)-\log(\int_{\Theta} \exp\{nL_n(t\}\,d\,t)
\, ,
\]
where $nL_n(t)=[\sum_{i=1}^n S(x_i)+\alpha]^{T}t-(n+\beta) A(t)\, .$
The minimum of $L_n(t)$ is achieved at
$$\tilde{\theta}_n=\tilde{\theta}(x^{(n)})=\dot {A}^{-1} (\frac{\sum_{i=1}^n S(x_i)+\alpha}{n+\beta})\, .$$
Notice that
$$\dot {A}(\tilde{\theta}_n)=\frac{\sum_{i=1}^n S(x_i)+\alpha}{n+\beta}=\frac{\sum_{i=1}^n S(x_i)}{n}+O(\frac{1}{n})=
\dot {A}(\hat{\theta}_n)+O(\frac{1}{n})\, .
$$
Through Taylor's expansion, it can be shown that
\begin{equation}\label{theta-approximation}
\tilde{\theta}_n=\hat{\theta}_n+O(\frac{1}{n})\, .
\end{equation}
Notice that
$
-\ddot{L}_n(t)=\frac{n+\beta}{n}\ddot {A}(t)\, .
$
Let
$
\Sigma=-\ddot{L}_n^{-1}(\tilde{\theta}_n)=\frac{n}{n+\beta}
\ddot {A}^{-1}(\tilde{\theta}_n)
\, .
$
By expanding $L_n(t)$ at the saddle point $\tilde{\theta}_n$
and applying the Laplace method (see \cite{DeBruijn-book}),
we have
\begin{eqnarray*}
\log(\int_{\Theta} \exp\{nL_n(t)\}\,d\,t)
&=&-\frac{d}{2}\log \, \frac{n}{2\pi}
+\frac{1}{2}\log (det\, \Sigma)+
nL_n(\tilde{\theta}_n)
+O(\frac{1}{n})\nonumber \\
&\longrightarrow&-\frac{d}{2}\log \, \frac{n}{2\pi}
-\frac{1}{2}\log (det\, I(\tilde{\theta}_n))+
nL_n(\tilde{\theta}_n) \, .
\end{eqnarray*}
Next,
\begin{eqnarray*}
& &{L_{mixture}}-nH({\hat \theta}_n)
\longrightarrow B(\alpha,\beta)+\frac{d}{2}\log \, \frac{n}{2\pi}+\frac{1}{2}\log (det\, I({\tilde \theta}_n))
-nL_n(\tilde{\theta}_n)-nH({\hat \theta}_n)
\\
&=& B(\alpha,\beta)+\frac{d}{2}\log \, \frac{n}{2\pi}+\frac{1}{2}\log (det\, I({\tilde \theta}_n))  \\
&&-[\sum_{i=1}^n S(x_i)+\alpha]^{T}{\tilde \theta}_n+(n+\beta) A({\tilde \theta}_n)
-n A(\hat{\theta}_n)+[\sum_{i=1}^n S(x_i)]^{T} {\hat{\theta}_n}  \\
&=&
\frac{d}{2}\log \, \frac{n}{2\pi}+\frac{1}{2}\log (det\, I({\tilde \theta}_n))
-[\alpha ^{T}{\tilde \theta}_n-\beta A({\tilde \theta}_n)- B(\alpha,\beta)]\\
&&-[\sum_{i=1}^n S(x_i)]^{T} (\tilde{\theta}-{\hat \theta}_n)+n[A({\tilde \theta}_n)-A({\hat \theta}_n)]
  \\
&=&
\frac{d}{2}\log \, \frac{n}{2\pi}+\frac{1}{2}\log (det\, I({\tilde \theta}_n))
-\log w({\tilde \theta}_n)\nonumber  \\
&& -[\sum_{i=1}^n S(x_i)^{T} (\tilde{\theta}-{\hat \theta}_n)+n\dot {A}(\hat{\theta})^{T}({\tilde \theta}_n-{\hat \theta}_n)
+\frac{n}{2}({\tilde \theta}_n-{\hat \theta}_n)^{T}\ddot A({\hat \theta}_n)
({\tilde \theta}_n-{\hat \theta}_n)]+O(\frac{1}{n})
\\
&=&
\frac{d}{2}\log \, \frac{n}{2\pi}+\frac{1}{2}\log (det\, I({\tilde \theta}_n))
-\log w({\tilde \theta}_n)\nonumber  \\
&& -n[[\bar{S}_n-\dot {A}({\hat \theta}_n)]^{T} (\tilde{\theta}_n-{\hat \theta}_n)
+\frac{1}{2}({\tilde \theta}_n-{\hat \theta}_n)^{T}\ddot A({\hat \theta}_n)
({\tilde \theta}_n-{\hat \theta}_n)]+O(\frac{1}{n})
\\
&\longrightarrow&
\frac{d}{2}\log \, \frac{n}{2\pi}+\frac{1}{2}\log (det\, I({\hat \theta}_n))
-\log w({\hat \theta}_n)\, .
\end{eqnarray*}
The last step is valid because of $\hat{\theta}_n=\dot {A}^{-1}(\bar{S}_n)$ and (\ref{theta-approximation}).
This proves the case of the prior $u(\theta)$ in (\ref{conjugate}). Meanwhile, we got the expansion of (\ref{conjugate-marginal})
\begin{eqnarray}
&&m(x^{(n)})=\exp\{-L_{mixture}\}=\exp\{B(\sum_{i=1}^nS(x_i)+\alpha,n+\beta)-B(\alpha,\beta)\} \nonumber \\
&=&\exp\{-nH({\hat \theta}_n)-\frac{d}{2}\log \, \frac{n}{2\pi}-\frac{1}{2}\log (det\, I(\hat{\theta}_n))
+\log w({\hat \theta}_n)+o(1)\} \nonumber \\
&=& [w({\hat \theta}_n)+o(1)]\exp\{-nH({\hat \theta}_n)-\frac{d}{2}\log \, \frac{n}{2\pi}-\frac{1}{2}\log (det\, I(\hat{\theta}_n))\}\, . \label{marginal-approximation}
\end{eqnarray}
If the prior of $\theta$ takes
the form of a finite mixture of the conjugate distributions (\ref{conjugate}) as in the following
\begin{equation}\label{mixture-conjugate}
w(\theta)=\sum_{j=1}^{J}\lambda_j\,\exp\{\alpha_{j}'\,\theta\,-\beta_{j}\,A(\theta)-B(\alpha_{j},\beta_{j})\}=\sum_{j=1}^{J}\lambda_j\,u_j(\theta)\, ,
\end{equation}
where $\sum_{j=1}^{J}\lambda_j=1$, $0<\lambda_j<1, j=1,\cdots, J$. Then the marginal density is given by
\begin{eqnarray*}%\label{mixture-marginal}
&&m(x^{(n)})=
\sum_{j=1}^{J}\lambda_j\,\exp\{B(\sum_{i=1}^nT(x_i)+\alpha_j,n+\beta_j)-
B(\alpha_j,\beta_j)\}\\
&=&\sum_{j=1}^{J}\lambda_j [u_j({\hat \theta}_n)+o(1)]\exp\{-nH({\hat \theta}_n)-\frac{d}{2}\log \, \frac{n}{2\pi}-\frac{1}{2}\log (det\, I(\hat{\theta}))\}\\
&=&\exp\{-nH({\hat \theta}_n)-\frac{d}{2}\log \, \frac{n}{2\pi}-\frac{1}{2}\log (det\, I(\hat{\theta}_n))\}[\sum_{j=1}^{J}\lambda_j u_j({\hat \theta}_n)+o(1)]
\\
&=&exp\{-nH({\hat \theta}_n)-\frac{d}{2}\log \, \frac{n}{2\pi}-\frac{1}{2}\log (det\, I(\hat{\theta}_n))\}[w({\hat \theta}_n)+o(1)]\\
&=&\exp\{-nH({\hat \theta}_n)-\frac{d}{2}\log \, \frac{n}{2\pi}-\frac{1}{2}\log (det\, I(\hat{\theta}_n))+\log w({\hat \theta}_n)+o(1)\}\, .
\end{eqnarray*}
Each summand was approximated by (\ref{marginal-approximation}). This completes the proof because of  $L_{mixture}=-\log\{m(x^{(n)})\}$.
\section{Acknowledgement}
The author is grateful to  Prof. Bin Yu and Dr. Jorma Rissanen for their guidance in learning the topic.
This research is supported by the National Key Research and Development Program of China (2022YFA1004801),
the National Natural Science Foundation of China (Grant No. 32170679, 11871462, 91530105),
the National Center for Mathematics and Interdisciplinary Sciences of the Chinese Academy of Sciences, and
the Key Laboratory of Systems and Control of the CAS.
\bibliographystyle{plain}

\bibliography{reference}

\end{document}